\documentclass[aps,prl,showpacs,reprint]{revtex4-1}
\usepackage{bm,epsfig,graphics,amssymb,amsmath,subeqnarray,setspace,graphicx,amsthm,epstopdf,subfigure,color,txfonts}

\usepackage{setspace,natbib}

\usepackage{epsfig,graphics,amssymb,amsmath,color,setspace}
\usepackage{graphicx}
\usepackage{natbib}
\usepackage[sfdefault=cmbr]{isomath}

\usepackage{graphicx,color,bm}
\graphicspath{{converted_graphics/}}
\usepackage{graphicx}


\begin{document}

\title{Chiral edge fluctuations of colloidal membranes}
\author{Leroy L. Jia$^1$,  Mark J. Zakhary$^{2}$, Zvonimir Dogic$^2$, Robert A. Pelcovits$^{3}$, and Thomas R. Powers$^{4}$ }
\affiliation{$^1$Division of Applied Mathematics, Brown University, Providence, RI 02912, USA}
\affiliation{$^2$The Martin Fisher School of Physics, Brandeis University, 415 South St, Waltham, MA 02454, USA}
\affiliation{$^3$Department of Physics, Brown University, 182 Hope Street, Providence, RI 02912, USA}
\affiliation{$^4$School of Engineering and Department of Physics, Brown University, 182 Hope Street, Providence, RI 02912, USA}

\date{\today}

\begin{abstract} 
We study edge fluctuations of a flat colloidal membrane comprised of a monolayer of aligned filamentous viruses. Experiments reveal that a peak in the spectrum of the in-plane edge fluctuations arises for sufficiently strong virus chirality. Accounting for internal liquid crystalline degrees of freedom by the length, curvature, and geodesic torsion of the edge, we calculate the spectrum of the edge fluctuations. The theory quantitatively describes the experimental data, demonstrating that chirality couples in-plane and out-of-plane edge fluctuations to produce the peak.\end{abstract}

\maketitle

Surfaces that resist bending are ubiquitous in biophysics and soft 
matter physics. The physics of enclosed cellular membranes~\cite{helfrich1973,HelfrichDeuling1975}, 
organelles such as the endoplasmic reticulum~\cite{Terasaki_etal2013,GuvenHuberValencia2014}, 
synthetic vesicles~\cite{SeifertBerndlLipowsky1991}, polymersomes~\cite{DischerAhmed2006}, surfactant interfaces~\cite{MilnerWitten1988}, and microemulsions~\cite{DeGennesTaupin1982} is described by a simple model that accounts for the energy cost of bending
with an effective bending modulus $\kappa$~\cite{canham1970,helfrich1973}. 
Furthermore, experiments have provided quantitative insight into how the bending modulus of such 2D assemblages depends on the properties of the constituent molecules~\cite{EvansRawicz1990,Rawicz:2000aa}. However, for many 
processes, such as vesicle fusion in exocytosis, trafficking of proteins, and the resealing of plasma membranes, the free energy associated with an exposed edge plays an equally important role~\cite{ChernomordikKozlov2008}. In conventional membranes edges are associated with transient states that quickly disappear as the assemblage seals itself, 
making it difficult to experimentally study the properties of 
the edges. 

Colloidal membranes are unique 2D assemblages comprised of a single liquid-like layer monolayer of aligned rodlike viruses that are held together by osmotic pressure~\cite{BarryDogic2010,Barry_etal2009,Gibaud_etal2012,Zakhary_etal2014}.  Although they are a few hundred times thicker, colloidal monolayer membranes share many properties common with lipid bilayers, such as in-plane fluidity and resistance to bending. However, they also display distinctive properties, such as propensity to 
have exposed edges, as well as shapes with negative Gaussian curvature~\cite{Gibaud_etal2017}. In this letter, we use experiments and theory to study the edge fluctuations of large, mostly flat colloidal membranes. We use an effective theory that treats the internal liquid-crystalline degrees of freedom using geometric properties of the membrane edge. In-plane fluctuations are mainly determined by the edge tension and associated bending rigidity. Out-of-plane height fluctuations distort the membrane surface leading to formation of excess saddle-splay deformation, and are thus influenced by the Gaussian curvature modulus. We show that the intrinsic chirality of the membrane couples in-plane and out-of-plane fluctuations yielding a fluctuation spectrum with an anomalous peak, and that this peak reflects the instability of a flat disk to a shape with helical edges.  

\begin{figure}[h!]
\centering
\begin{minipage}{0.1\textwidth}
\begin{subfigure}[]
\centering
\includegraphics[scale = 0.4]{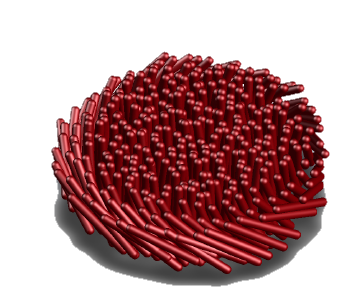}
\end{subfigure}\\
\begin{subfigure}[]
\centering
\includegraphics[scale = 0.4]{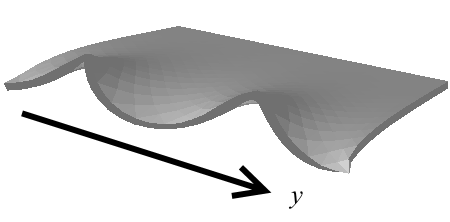}
\end{subfigure}
\end{minipage}
\hfill
\begin{minipage}{0.2\textwidth}
\begin{subfigure}[]
\centering
\includegraphics[scale = 0.48]{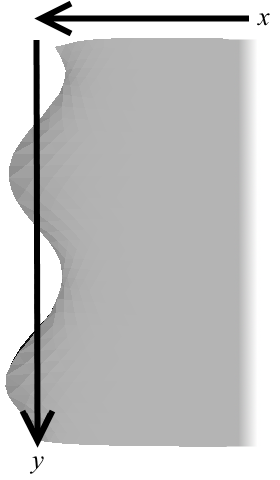}
\end{subfigure}
\end{minipage}
\caption{(Color online.) (a)  Schematic of a colloidal membrane (b) Semi-infinite minimal surface whose boundary is a helix with wavenumber $q$. 
(c) Same shape viewed from above, looking down the $z$-axis at the $x$-$y$ plane. }\label{edgefig}
\end{figure}

Colloidal membranes were assembled by mixing a dilute isotropic suspension of monodisperse rod-like {\it fd-wt} viruses  with a non-adsorbing polymer, Dextran (M.W. 500,000, 37\,mg/ml)~\cite{BarryDogic2010}. The \textit{fd-wt} filaments are 0.88\,$\mu$m long and have a diameter of $\approx6$\,nm. The rods are parallel to the surface normal and to each other in the membrane interior, but they twist at the edge [Fig.~\ref{edgefig}(a)] to minimize the interfacial area between the rods and the enveloping polymer depletant~\cite{Barry_etal2009,KangGibaudDogicLubensky2015}. Increasing the rod chirality raises the free energy of the interior untwisted rods and lowers the free energy of the twisted rods near the edge~\cite{Gibaud_etal2012}. 
Chirality of {\it fd-wt} increases with decreasing temperature~\cite{DogicFraden2000}, enabling  {\it in situ} control of the edge tension. With decreasing temperature the edge tension becomes sufficiently low that a 2D membrane becomes unstable and undergoes a transition into 1D twisted ribbons~\cite{Gibaud_etal2012}. 

\begin{figure}[h!]
  \centering
        \begin{subfigure}[][]
        \centering
        \includegraphics[scale=0.32]{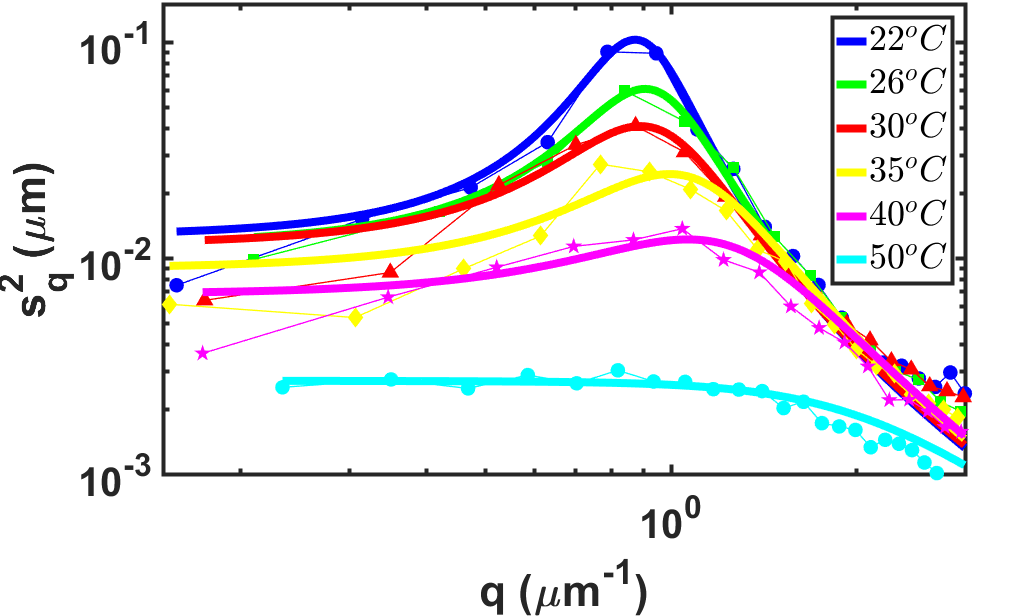}
    \end{subfigure}
    \begin{subfigure}[]
        \centering
        \includegraphics[scale=0.45]{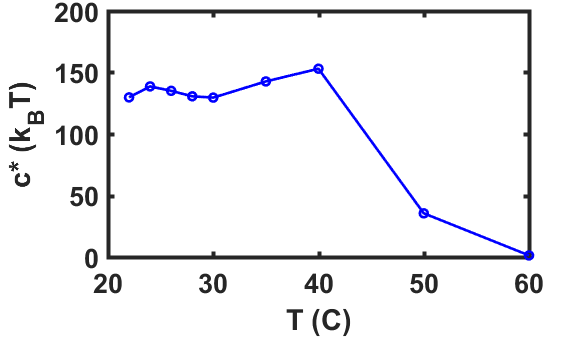}
    \end{subfigure}
    \begin{subfigure}[]
        \centering
        \includegraphics[scale=0.45]{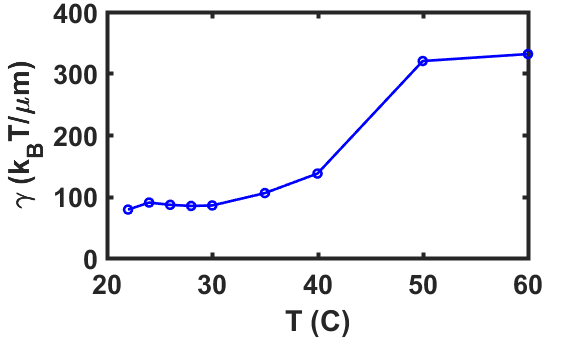}
    \end{subfigure}
    \caption{(Color.) (a) Normalized power spectrum  $s_q^2\equiv q^2\langle u_q{u_{-q}}\rangle$ in which $c^*$ (in units of $k_BT$) and $\gamma$ (in units of $k_BT/\mu$m) are fit to the data, with $\kappa/\bar{\kappa}\rightarrow\infty$, $B=100k_\mathrm{B}T\mu$m, and $\bar{\kappa}=150k_\mathrm{B}T$. The value of $B$ used was found by fitting the $T=60^\circ$\,C data to the achiral formula, Eq.~(\ref{oldps}). (b) Values of $c^*$ obtained from fitting. (c) Values of $\gamma$ obtained from fitting.}\label{powerfigth}
\end{figure}

Following previously published methods~\cite{Zakhary_et.al_SoftMatter2013}, we measured the in-plane fluctuation spectrum of an exposed colloidal membrane edge (see Supplementary Material~\footnote{See Supplemental Material at [URL inserted by publisher] for movies of edge fluctuations at different temperatures.}). The edge fluctuations were quantified over a range of temperatures (Fig.~\ref{powerfigth}). For all conditions, the curves tend to a constant value at small wavenumber $q$ and fall off as $1/q^2$ at large wavenumber, as in the previous measurements. However, for strongly chiral systems  at lower temperatures, a peak develops around $q=1$\,$\mu$m$^{-1}$. We note that the measured fluctuation spectrum depends on the purity of the virus preparation as well as the depleting polymer Dextran. Certain virus preparations do not exhibit the membrane-to-ribbon transition. These samples also do not have a fluctuation spectrum with a well-defined peak. The exact nature of the 
contaminants in these samples has not been determined. Previous work has demonstrated that even a single actin filament exhibits a strong tendency to dissolve at the edge, and can suppress the polymorphic transition~\cite{Zakhary_et.al_SoftMatter2013}. In the remainder of this work we restrict our analysis to only those sample preparations that exhibited a well-defined ribbon-to-membrane transition and thus the anomalous peak.

The in-plane edge fluctuations in the high-temperature achiral limit are described by a simple model, which approximates a flat, circular membrane as a semi-infinite membrane with an infinite straight edge~\cite{Gibaud_etal2012}. 
For in-plane fluctuations, the effective energy of the edge is given by
\begin{equation}
E_1=\int\mathrm{d}s\left[\gamma+\frac{B}{2}k^2\right],\label{Eedge}
\end{equation}
where  $s$ is the edge arclength, $\gamma$ is the line tension, $B$ is the bending stiffness, and $k$ is the curvature. For a flat membrane lying in the $x$-$y$ plane, we describe the path of the edge by $(u(y),y,0)$, where $u(y)$ measures the local deviation of the edge from being perfectly straight. Expanding the energy~(\ref{Eedge}) to second order in $u$ and applying the equipartition theorem yields
\begin{equation}
\langle u_q u_{-q}\rangle=\frac{k_\mathrm{B}T}{B q^4+\gamma q^2},\label{oldps}
\end{equation}
where $u_q$ is the Fourier amplitude defined by 
$u(y) =({1}/{\sqrt{L}} )\sum_{q} u_q \exp{\mathrm{i}qy}$,
and we have enforced periodic boundary conditions with period $L$. Fitting the high-temperature spectrum of achiral rods to Eq.~(\ref{oldps}) yields $\gamma$ and $B$~\cite{Gibaud_etal2012}.

As mentioned above, the line tension $\gamma$ depends on chirality. We can estimate this dependence by examining the energy density of the twisted rods near the edge of a semi-infinite membrane~\cite{Barry_etal2009}. The total energy per unit length of a flat membrane in the limit of weak chirality is $\gamma=\gamma_0-K_2\lambda q_0^2/2$, where $\gamma_0$ is the line tension of an achiral membrane, $K_2$ is the twist Frank constant, $\lambda$ is the twist penetration depth, and $q_0$ is the preferred twist. The bend modulus $B$ may also be estimated by considering the liquid crystal degrees of freedom at the edge. 
The edge of colloidal membrane assumes a surface-tension-minimizing semi-circular profile. Consequently, at the edge, the rod-like viruses lie in the plane of the membrane. If the edge is curved, these rods will not be aligned, and will thus give rise to a liquid-crystal bend energy penalty. 
Using the twist penetration depth $\lambda$ and the Frank elastic constant $K_3$ for bend, we find $B\approx D K_3\lambda$, where $D\approx1\,\mu$m is the membrane thickness.
Assuming $K_3\approx100$\,$k_\mathrm{B}T/\mu$m and $\lambda\approx1\,\mu$m, yields $B\approx100$\,$k_\mathrm{B}T\,\mu$m, which agrees with measurements of the in-plane fluctuations of the edge of a large flat membrane ~\cite{Gibaud_etal2012}. 

Next we show that chirality  also couples the in-plane and out-of-plane fluctuations of the edge, and ultimately leads to a different $q$-dependence in the power spectrum if the chirality is sufficiently strong.
Using the full liquid crystal theory for colloidal membranes~\cite{KaplanTuPelcovitsMeyer2010} to calculate the  power spectrum of the fluctuating edge of a curved surface is a daunting task. We note that the twist of the virus particles is limited to a region near the edge, which is much thinner than the membrane size. Therefore, we use an effective theory in which the energetic cost of the nonuniform liquid crystalline distortions are described by the membrane's geometric properties alone. A closely related effective theory has been successfully used to calculate the scalloped shapes of colloidal membranes composed of a mixture of viruses of opposite handedness~\cite{KaplanGibaudMeyer2013,Gibaud_etal2017}. The edge energy $E_1$ (\ref{Eedge}) is invariant under mirror reflections and is therefore achiral. To account for the chirality of the membrane, in addition to the tension and bending terms, we include a mirror-symmetry-breaking term in the edge energy~\cite{Zhong-canJixing1991,TuSeifert2007,KaplanGibaudMeyer2013}:
\begin{equation}
E^*=c^*\int\mathrm{d}s\tau_\mathrm{g}.\label{Echiral}
\end{equation}
The geodesic torsion $\tau_\mathrm{g}$ is the rate that the surface normal $\hat{\mathbf{n}}_C$  at the edge $C$  twists about the tangent vector $\hat{\mathbf{T}}$~\cite{struik1988}: 
 \begin{equation}
 \tau_\mathrm{g}=\hat{\mathbf{T}}\cdot\hat{\mathbf{n}}_C\times\mathrm{d}\hat{\mathbf{n}}_C/\mathrm{d}s.
 \end{equation} 
Note that unlike the ordinary torsion of a curve~\cite{struik1988}, the geodesic torsion is well-defined even for a straight line. Furthermore, the term $E^*$ is invariant under replacing the surface normal by its reverse, $\hat{\mathbf{n}}\mapsto-\hat{\mathbf{n}}$, as expected for a symmetric membrane. Also, $E^*$ breaks mirror symmetry because the sign of the geodesic torsion changes when the handedness of the edge changes. Since  the preference for a definite handedness of the edge ultimately arises from the intrinsic twist of the virus particles, the modulus $c^*$ must be proportional to $q_0$, the preferred rate of twist of the viruses. We estimate $c^*\approx DK_2 q_0 \lambda$, where $K_2$ is the twist elastic constant. For $q_0\approx 1\,\mu$m$^{-1}$ and $K_2\approx K_3$, $c^*\approx100\,k_\mathrm{B}T$.

The absence of mirror symmetry leads to a preference for helical edge fluctuations that  couple in-plane and out-of-plane fluctuations. 
Since the distortions of the membrane with a helical edge penetrate into the interior, we must also consider the membrane bending energy. For a thin membrane, the bending energy is given by the Canham-Helfrich energy~\cite{canham1970,helfrich1973},
\begin{equation}
E_{2}=\frac{\kappa}{2}\int\mathrm{d}A (2H)^2+\bar{\kappa}\int\mathrm{d}A K,
\end{equation}
where $H=(1/R_1+1/R_2)/2$ is the mean curvature, $R_1$ and $R_2$ are the principal radii of curvature, $K=1/(R_1R_2)$ is the Gaussian curvature, $\kappa$ is the bending modulus, and $\bar{\kappa}$ is the Gaussian curvature modulus. Although the thickness of the membrane, $D\approx1\,\mu$m, is comparable to the length scale $q^{-1}\approx 1\,\mu$m of ripples observed at the edge of a  membrane disk undergoing the transition to a twisted ribbon,  we will proceed with the assumption that the membrane is thin. Note that since the membrane has an edge, the Gauss-Bonnet theorem implies that the contribution from the bending energy from the Gaussian curvature term depends on membrane shape, in contrast with the case of a closed vesicle~\cite{struik1988,kamien2002}.

Experimental observations suggest that $\kappa$ is very large. The balance between membrane bending energy and edge energy cost suggests that above a critical size of $2\kappa/\gamma$~\cite{Helfrich1974,boal2002} (disregarding $\bar{\kappa}$ for now), a  flat membrane should transform into a closed vesicle. For conventional lipid bilayers $\kappa\approx 10^{-19}\,$J and $\gamma\approx10^{-11}\,$J/m~\cite{Evans_etal2003}, leading to a critical length scale of $R\approx20$\,nm.
We have not observed 3D colloidal membrane vesicles, yet routinely observe flat colloidal membranes with $R\approx100\,\mu$m, indicating that for the typical scale of colloidal membrane line tension, $\gamma\approx100\,k_\mathrm{B}T\,\mu$m, the bending modulus $\kappa$ of colloidal membranes is larger than $\approx5000\,k_\mathrm{B}T$. Furthermore, colloidal membranes commonly assume the shapes of minimal surfaces ($H=0$), such as twisted ribbons, which also suggests a large value of $\kappa$. Recent measurements and theoretical estimates of the Gaussian curvature modulus show that $\bar{\kappa}\approx 200\,k_\mathrm{B}T$ in colloidal membranes~\cite{Gibaud_etal2017}. Note that $\bar{\kappa}>0$; for lipid bilayer membranes $\bar{\kappa}$ is typically negative due to the compressive stress in the head groups~\cite{SiegelKozlov2004,HuBriguglioDeserno2012,Hu_etal2013}, although it  can be positive in smectic liquid crystals~\cite{BoltenhagenLavrentovichKleman1992} and block copolymers~\cite{Wang1992}. For the remainder of this article we assume $\kappa\gg\bar{\kappa}$. 

%
To study the stability of a flat membrane and the out-of-plane fluctuations of its edge, we calculate the energy of a semi-infinite membrane with a rippled edge, working to second order in the deformation. The membrane is initially flat with mid-surface at the $z=0$ plane and occupying $x<0$. The $y$-axis is the initially straight edge. We perturb the surface by deforming the edge so that  the position of points on the membrane above the coordinates in the plane $(x,y)$ are given by $\mathbf{R}(x,y)=(x,y,\zeta(x,y))$, and with the edge given by  $\mathbf{R}_C(y)=(u(y),y,v(y))$, which implies the boundary condition $\zeta(u(y),y)=v(y)$. If $u(y)$ and $v(y)$ are sinusoidal and out of phase, the edge will have a helical nature {as illustrated in Fig. \ref{edgefig}(b,c)}. Care must be taken in calculating the edge quantities since we must expand both the quantities themselves and their arguments. For example, to find the geodesic torsion at the edge we expand the argument of the normal at the edge, 
$
\mathbf{n}_C(u(y),y)\approx \mathbf{n}(0,y)+u{\partial\mathbf{n}_C}/{\partial x}|_{x=0}$.

Expanding the energy to second order, we find the total energy $E=E_1+E_2+E^*$,
\begin{eqnarray}
E&=&\int\mathrm{d}x\mathrm{d}y\left\{\frac{\kappa}{2}(\nabla^2\zeta)^2+\bar{\kappa}\left[\frac{\partial^2\zeta}{\partial x^2}\frac{\partial^2\zeta}{\partial y^2}-\left(\frac{\partial^2\zeta}{\partial x\partial y}\right)^2\right]\right\}\nonumber\\
&+&\int\mathrm{d}y\left[\frac{\gamma}{2}\left(u'^2+v'^2\right)+\frac{B}{2}\left(u''^2+v''^2\right)+c^*u'v''\right],\label{energy2}
\end{eqnarray}
where the prime denotes the derivative with respect to $y$, e.g. $u'=\mathrm{d}u/\mathrm{d}y$. 
To first order the Euler-Lagrange equation for the energy~(\ref{energy2}) is
$(\nabla^2)^2\zeta=0$.
Since the horizontal and vertical positions of the edge are prescribed, we do not enforce the force boundary conditions at the edge. The condition of zero bending moment at the boundary  to first order~\cite{landau_lifshitz_elas} is $\kappa\nabla^2\zeta+\bar{\kappa}{\partial^2\zeta}/{\partial y^2}=0$.
The solution to the Euler-Lagrange equations that satisfies this boundary condition and {$\zeta(u(y),y)=v(y)$} in the limit of large bending stiffness, $\kappa\gg\bar{\kappa}$, is a minimal surface,
$\zeta(x,y)={1}/{\sqrt{L}}\sum_qv_q\exp\left(|q|x+\mathrm{i}qy\right)$.
To keep the area fixed, the whole surface must shift in the negative $x$-direction; however, this shift is second order in the $v_q$ and does not affect the energy to leading order. 

Inserting the Fourier expansions of $\zeta(x,y)$, $u(y)$, and $v(y)$ into the total energy yields
\begin{eqnarray}
E=\sum_q&&\left[\left(-\bar{\kappa}|q|^3+\frac{\gamma}{2}q^2+\frac{B}{2}q^4\right)|v_q|^2\right.\nonumber\\
&+&\left.\left(\frac{\gamma}{2}q^2+\frac{B}{2}q^4\right)|u_q|^2
-\frac{\mathrm{i}}{2}c^*q^3 \left(u_q v_{-q}-u_{-q}v_q\right)\right].~\label{Eq}
\end{eqnarray}
Using the expression for the energy~(\ref{Eq}) we can study the stability of a flat semi-infinite membrane. We first consider the achiral case, $c^*=0$, which applies at high temperatures ($T\gtrsim60^\circ$\,C). The horizontal and vertical fluctuations of the edge are decoupled, and ripples $u_q$ in the plane of the membrane always increase the energy. However, when the Gaussian curvature modulus is large enough, $\bar{\kappa}=\sqrt{\gamma B}\approx 100k_\mathrm{B}T$, ripples $v_q$  in the vertical (positive $z$-) direction with wavenumber with $q_\mathrm{c}=\bar{\kappa}/B\approx 1\,\mu$m$^{-1}$  decrease the energy. 

In the chiral case, we study the stability of the flat membrane with an edge by writing $E=\sum_q (u_q,v_q)M_q(u_{-q},v_{-q})^\mathrm{T}/2$ and diagonalizing $M_q$ to find its eigenvalues
$
\sigma_{\pm}=\gamma q^2+Bq^4-|q|^3\left(\bar{\kappa}\mp\sqrt{c^{*2}+\bar{\kappa}^2}\right)$.
In this case, either chirality or the Gaussian curvature modulus can drive an instability. The condition for a rippled edge is
\begin{equation}
\bar{\kappa}+\sqrt{c^{*2}+\bar{\kappa}^2}\ge2\sqrt{\gamma B}.\label{critkc}
\end{equation}
When the condition~(\ref{critkc}) is just satisfied, the unstable wavenumber is 
$q_\mathrm{c}=\left({\bar{\kappa}+\sqrt{c^{*2}+\bar{\kappa}^2}}\right)/({2B})$.
Note that the factor of $\mathrm{i}c^*$ in the off-diagonal components of $M_q$ leads to a phase difference between the $x$- and $z$- components of the eigenvectors of $M_q$, and thus a preferred handedness to the edge depending on the sign of $c^*$.

The energy~(\ref{Eq}) along with the equipartition theorem yields the power spectrum for in-plane and out-of-plane edge fluctuations, which is valid for stable flat states
\begin{eqnarray}
\langle u_q u_{-q}\rangle&=\dfrac{k_\mathrm{B}T}{Bq^4+\gamma q^2-2c^{*2}q^4/[\bar\kappa q  - 2 (Bq^2 + \gamma)]}
\label{uquqth}\\
\langle v_q v_{-q}\rangle&=\dfrac{k_\mathrm{B}T}{Bq^4+\gamma q^2 -\frac{1}{2}\bar\kappa |q|^3 -c^{*2}q^4/(Bq^2 + \gamma)}.
\end{eqnarray}

Note that $c^*$ and $\bar{\kappa}$ only affect the fluctuations for intermediate $q$; the large and small $q$ behavior is controlled by the bending stiffness and line tension, respectively, just as in the case of an achiral membrane. Note also that the out-of-plane fluctuations of the edge of an achiral membrane have a distinctly different $q$-dependence than the in-plane fluctuations of Eq.~(\ref{oldps}): $\langle v_qv_{-q}\rangle =k_\mathrm{B}T/(Bq^4-\frac{1}{2}\bar{\kappa}|q|^3+\gamma q^2)$ when $c^*=0$.

As the temperature is lowered the system comes closer to fulfilling the condition~(\ref{critkc}) for ripples to form,  and a peak appears near $q_\mathrm{c}$. Figure~\ref{powerfigth} shows fits of the theoretical expression~(\ref{uquqth}) for $s_q^2\equiv q^2\langle u_q{u_{-q}}\rangle$ assuming $\kappa/\bar{\kappa}\rightarrow\infty$, $B=100\,k_\mathrm{B}T\,\mu$m, and $\bar{\kappa}=150\,k_\mathrm{B}T$, along with the values of $\gamma$ and $c^*$ obtained from fitting each curve. The value of $B$ used was found by fitting the $T=60^\circ$\,C data to the achiral formula, Eq.~(\ref{oldps}). Because $B$ does not change appreciably with $T$, all curves collapse onto a single line in the large $q$ limit. Similarly, $\bar\kappa$ is not expected to depend significantly on $T$ and was fixed for fitting. Because both $\bar\kappa$ and $c^*$ control the size of the peak, fixing $\bar\kappa$ also allows the effect of $c^*$ to be assessed more accurately. The magnitude of the $\bar{\kappa}$ used in our fits is comparable with the recent experimental measurements and theoretical estimates of $\bar{\kappa}$~\cite{Gibaud_etal2017}.

The values of $\gamma$ and $c^*$ from the fitting
have the expected order of magnitude and obey the expected trend of $\gamma$ increasing and $c^*$ decreasing to zero as the temperature increases. Although the fits capture the shape of the peak well, there is some discrepancy with the experimental data at the smallest measured values of $q$. There are two main reasons for this discrepancy. First, the characteristic widths of the peaks are fairly large, and there are not enough data points taken at small enough $q$ to escape the influence of the peak. Second, when  $q$ decreases, the fluctuation relaxation time increases rapidly. The longer relaxation time leads to poor statistics, since it reduces the number of configurations over which the data can be averaged. Consequently, the fits tend to underestimate $\gamma$ when the temperature is low. 

Early work demonstrated that the height fluctuations of colloidal membranes scaled as $1/q^3$~\cite{BarryDogic2010}, leading to an estimate of $\kappa \approx150 k_BT$, in contrast with our arguments above for a much larger value of $\kappa$. The $1/q^3$ scaling was attributed to the fact that the analysis only accounted for 1D cut of a 2D membrane that is viewed in the edge-on configuration~\cite{MutzHelfrich1990}. When combined with the appropriate line tension, $\kappa$ extracted from this measurement indicates that all colloidal membranes should fold into vesicles, which is not observed experimentally. Our analysis of edge fluctuations may provide a plausible resolution to this contradiction. In the original experiments~\cite{BarryDogic2010} the distance from the membrane edge was not controlled. It seems possible that the original work analyzed the much softer edge fluctuations, instead of fluctuations associated with the bending of the membrane interior. 
To distinguish between these two options, it will be important to measure how fluctuations depend on the distance away from the edge. 

To conclude, we have measured the small-amplitude fluctuations of the edge of a colloidal membrane, and found that as the chirality increases, a peak forms at a characteristic wavelength. Our effective geometric theory captures the important features of the measurement such as the formation of the peak, and shows how the Gaussian curvature modulus affects the fluctuations when chirality couples the undulations of the edge in and out of the plane of the membrane. We have also calculated the power spectrum for out-of-plane fluctuations of the edge, which would be especially interesting to measure in the achiral case, as it offers another methods of estimating the Gaussian curvature modulus $\bar{\kappa}$. 

TRP thanks Brandeis University for its hospitality while some of this work was done. We thank Andrew Balchunas, Art Evans, and Jemal Guven for helpful discussions.  This work was supported in part by the National Science Foundation  through grants MRSEC-1420382 (ZD, RAP, and TRP), BMAT-1609742 (ZD)  and CMMI-1634552 (RAP and TRP).


%

%
%
%

\end{document}